\documentclass[reprint,amsmath,amssymb,aps,showpacs]{revtex4-1}
\usepackage{epsfig}
\usepackage{dcolumn}
\usepackage{bm}
\graphicspath{{figs/}} 
\begin{document}
\title{Measurement of the Shape Factor for the Beta Decay of $^{14}$O}
\author{E.A.\ George}
\affiliation{Physics Department, Wittenberg University, 
Springfield Ohio 45501, USA}
\author{P.A.\ Voytas}
\affiliation{Physics Department, Wittenberg University, 
Springfield Ohio 45501, USA}
\author{G.W.\ Severin}
\altaffiliation{Present address: The Hevesy Laboratory, Center 
\affiliation{Physics Department, University of Wisconsin-Madison, Madison,
Wisconsin 53706, USA}
for Nuclear Technologies, Technical University of Denmark, 
Frederiksborgvej 399, 4000 Roskilde, Denmark.}
\author{L.D.\ Knutson}
\email{Email contact: knutson@physics.wisc.edu}
\affiliation{Physics Department, University of Wisconsin-Madison, Madison,
Wisconsin 53706, USA}
\date{\today}
\begin{abstract}
We report results from an experiment designed to test the
conserved vector current (CVC) hypothesis by measuring
the shape of the $\beta$-decay spectrum for the 
allowed $0^+\rightarrow 1^+$ ground state decay
of $^{14}$O.  Measurements of the spectrum intensity were obtained
with a superconducting beta spectrometer and will be reported
for positron kinetic energies ranging from 1.9 to 4.0 MeV.
After dividing out phase space, Coulomb, and other correction factors, the
resulting shape function has a negative slope of
several per cent per MeV.  We define a parameter $a'$,
which is essentially a measure of the average slope of
the shape function over the energy range of the measurements, and determine
its value to be $a' = -0.0290 \pm 0.0008$ (stat.) $\pm 0.0006$ (syst.).
The measured slope parameter is in good agreement with predictions from
shell model calculations that respect CVC.
\end{abstract}
\maketitle

\section{Introduction}
\label{Intro}

The nucleus $^{14}$O has a half-life of
70.62\kern 2pt s \cite {thalf} and decays by positron emission.
More than 99\% of the decays proceed by a $0^+\rightarrow 0^+$
Fermi transition from the $^{14}$O ground
state to the isobaric analog 2.313 MeV first excited
state of $^{14}$N.  The ground state branch, which is the subject of the
present work, is an allowed $0^+\rightarrow 1^+$
Gamow-Teller (GT) transition with an 
endpoint energy of 4.12 MeV, and a 
log\kern 1pt $ft$ value of roughly 7.3.  
The unusually large $ft$ is
thought to be the result of an accidental cancellation between various
nuclear wave function components that contribute to the axial vector
matrix element \cite{I53,V57}.

Because the allowed GT matrix element, 
$\langle \sigma \rangle$,
is suppressed, contributions from
ordinarily small forbidden matrix elements may well be appreciable, 
and could lead to deviations of the beta spectrum from the purely
statistical shape.  

Of particular interest is the contribution from the (vector) weak magnetism 
(WM) term.  The WM matrix 
element
affects the spectrum shape through interference with the dominant
GT matrix element (see for example Ref.\ \cite {CH})
giving rise to an extra energy dependent shape factor,
\begin{equation}
S_0(E) \simeq 
\left[1 - {4\over 3M}{\langle {\rm WM}\rangle \over \langle 
\sigma \rangle}  
\left(E - {E_0\over 2} - {m_e^2\over 2E} \right)\right],
\label{eq:cvc}
\end{equation}
where $M$ ($m_e$) is the nucleon (electron) rest energy, 
$E$ is the total electron energy, 
and $E_0$ is the corresponding endpoint energy. The quantity
$\langle {\rm WM}\rangle = b/A$ is the WM matrix element,
where $b$ is defined in Ref.\ \cite{CH} and $A=14$.

In 1958, Gell-Mann  \cite {G-M} proposed that measurements of 
$\langle {\rm WM}\rangle$ in systems like the present one can, in principle,
allow a test of the conserved vector current (CVC) hypothesis \cite {CVC}.
CVC implies that the WM operator is identical to the
electromagnetic M1 operator
that determines the lifetime of the 2.313 MeV 
state in $^{14}$N.  Assuming
CVC and charge symmetry, one would have
$\langle {\rm M1}\rangle = \langle {\rm WM}\rangle$, with 
$\langle {\rm M1}\rangle$ known from the measured $\gamma$-decay width.

Following Gell-Mann's suggestion, a number of experimental groups 
\cite {A12,B12,C12,D12} 
undertook
experiments to measure $\beta$-decay shape factors in the $A=12$ system.
Unfortunately, results from the different groups show discrepancies well beyond
the quoted uncertainties, demonstrating the extreme difficulty
of experiments of this kind.

Calaprice and Holstein \cite {CH} have calculated $\langle {\rm WM} \rangle$
for a series of nuclei and found that the ratio 
$\langle {\rm WM} \rangle/\langle \sigma \rangle $ should be 
an order of magnitude larger for $^{14}$O than for the 
$A=12$ nuclei. From the experimental point of view, this makes 
the $^{14}$O experiment attractive.
On the other hand, given that $\langle \sigma \rangle $ is so small,
one needs to be concerned about possible energy dependences that can
arise from the various higher order matrix elements (for example, second
forbidden terms) or from other 
normally negligible effects such as charge symmetry violation.

Tests of CVC are of central importance and this provides the fundamental
motivation for the present experiment.  According to CVC, the 
weak charge-changing vector currents together with the electromagnetic
current make up a 3 component isospin multiplet.  This symmetry
leads to the $\langle {\rm WM} \rangle = \langle {\rm M1}  \rangle$
result.  CVC has other consequences as well, such as the non-renormalization
of the weak vector current, but 
$\langle {\rm WM} \rangle = \langle {\rm M1}  \rangle$ experiments
are considered strong tests of CVC \cite{grenacs}. 
Previous $^{14}$O measurements (from the mid 1960's) have been reported
\cite {sidhu}, but in view of the importance of the subject we believe that
a second measurement of the spectrum shape would be valuable, particularly
since recent analyses have suggested that there may be
systematic problems with these measurements.

Measurements of the $\beta $-spectrum of $^{14}$O are important for a second
reason.  The excited state decay is one of the $0^+\rightarrow 0^+$
superallowed Fermi transitions used to determine the $V_{ud}$ element
of the CKM matrix, and the analysis requires knowledge of the
$^{14}$O branching ratio.  We will report new results for that quantity
in a subsequent publication.

\section{Summary of Previous Work}
\label{History}
 
Measurements of the shape of the $\beta$ spectrum of $^{14}$O were reported in
1966 by Sidhu and Gerhart \cite{sidhu} (SG).  These authors used an iron-free, 
uniform-field, solenoidal spectrometer to focus positrons emitted
from a source that was produced by freezing $^{14}$O water onto a 
liquid-nitrogen-cooled beryllium disc.  Positrons passing
through the spectrometer  were detected with a plastic scintillator 
approximately 1 cm thick. Corrections for backscattering
from the source backing, for $\gamma$-ray backgrounds, and for sub-threshold
positron events were included in the data analysis.
 
The experimental results summarized in Figure 6 of Ref.\ \cite {sidhu}
show significant deviations from the purely allowed (statistical) shape.
The authors plot the quantity
\begin{equation}
S(E) = {N\over p^3(E - E_0)^2 F_0(p,Z)},
\end{equation}
where $N$ is the number of detected positron events (normalized
for variations in source activity and counting time),
and where $F_0$ is the Fermi function.
The ``extra'' factor of $p$ in the denominator
is included to account for the fact
that the momentum acceptance width of the spectrometer 
scales with $p$.

The experiment shows $S(E)$ to be a monotonically decreasing quantity
with a relative slope of typically 9\% per MeV;
more specifically, if we compare the reported measurements
with a function of the form given in Eq. (\ref {eq:cvc}),
\begin{equation}
 S(E) \sim 1 - \alpha \left(E - {E_0\over 2} - {m_e^2\over 2E} \right),
\label{eq:slope}
\end{equation}
the results indicate a slope parameter $\alpha$ of about 0.09/MeV.
This is roughly a factor of 2 larger than the naive CVC prediction
from Ref.\ \cite{CH}.  However, it should be noted that this comparison
does not take into account various modern corrections to the beta
spectrum shape.  
In particular, as we shall discuss later,
corrections for radiative processes are not negligible.

On the theoretical side, Garc{\'i}a and Brown \cite {GB} (GB) 
have carried out a detailed study
of the $A=14$ $\beta$-spectrum shapes and $ft$ values.  One of the 
long-standing issues in mass 14 is the large asymmetry in the 
$ft$ values for the $^{14}$C and $^{14}$O decays.
It is thought that the cancellations in the dominant GT matrix element
may be very sensitive to small wave function differences that can arise
from charge symmetry violations and/or Coulomb effects.
GB investigate whether these same effects may also
be responsible for the unexpectedly large $^{14}$O $\beta$-decay
slope parameter.  They conclude that
these effects can be no more than a few per cent for
 $\langle {\rm WM}\rangle$ and $\langle {\rm M1}\rangle$.
The slopes they find in calculations that respect
charge symmetry and CVC are at
least a factor of 1.7 smaller than the measured slope.
 
Towner and Hardy \cite {TH} (TH) have reported a 
new analysis of the $^{14}$O $\beta$-decay
data.  They, for the first time, apply corrections
for radiative processes as well as several other small effects.
Nevertheless, they still agree with the general conclusions of
GB when free-nucleon operators are used in their calculations.
Attempts to reproduce the SG data seem to require violation of CVC.

 On the other hand, when renormalized operators
are used for the GT and WM matrix elements, the results
improve significantly.
It is known \cite {raman} that, in finite nuclei, the effective axial vector
coupling constant is depressed (compared
to the free nucleon value) by core polarization and
meson exchange currents, and it is expected that 
the M1 and WM operators also need to be renormalized.  
TH use the known GT renormalization factor
along with M1 
renormalization parameters from Ref.\ \cite {khanna}, and fit the
$\beta $ decay data of Ref.\ \cite {sidhu}
with only a single adjustable wave function parameter.  
In doing so, they are able to reproduce the overall 
transition rate and obtain a $\langle {\rm WM}\rangle$ value
consistent with CVC, while underpredicting the measured slope
by only around 20\% instead of by almost a factor of two.

As we suggested earlier, contributions from higher order
matrix elements may be of importance 
in $^{14}$O $\beta$ decay.  GB include two higher order terms in
in their analysis, and it appears to us (from calculations
based on formulas presented by GB) that these
terms have a significant effect on the slope parameter.
TH do not explicitly separate out the higher order
pieces, but their results also suggest that these terms are important. 
In particular, their
calculations predict the presence of an 
$E^2$ term in the shape-correction function which is
an order of magnitude larger than one would obtain if
only the GT and WM terms are present.

\begin{figure*}
\centerline{\includegraphics[width=140mm]{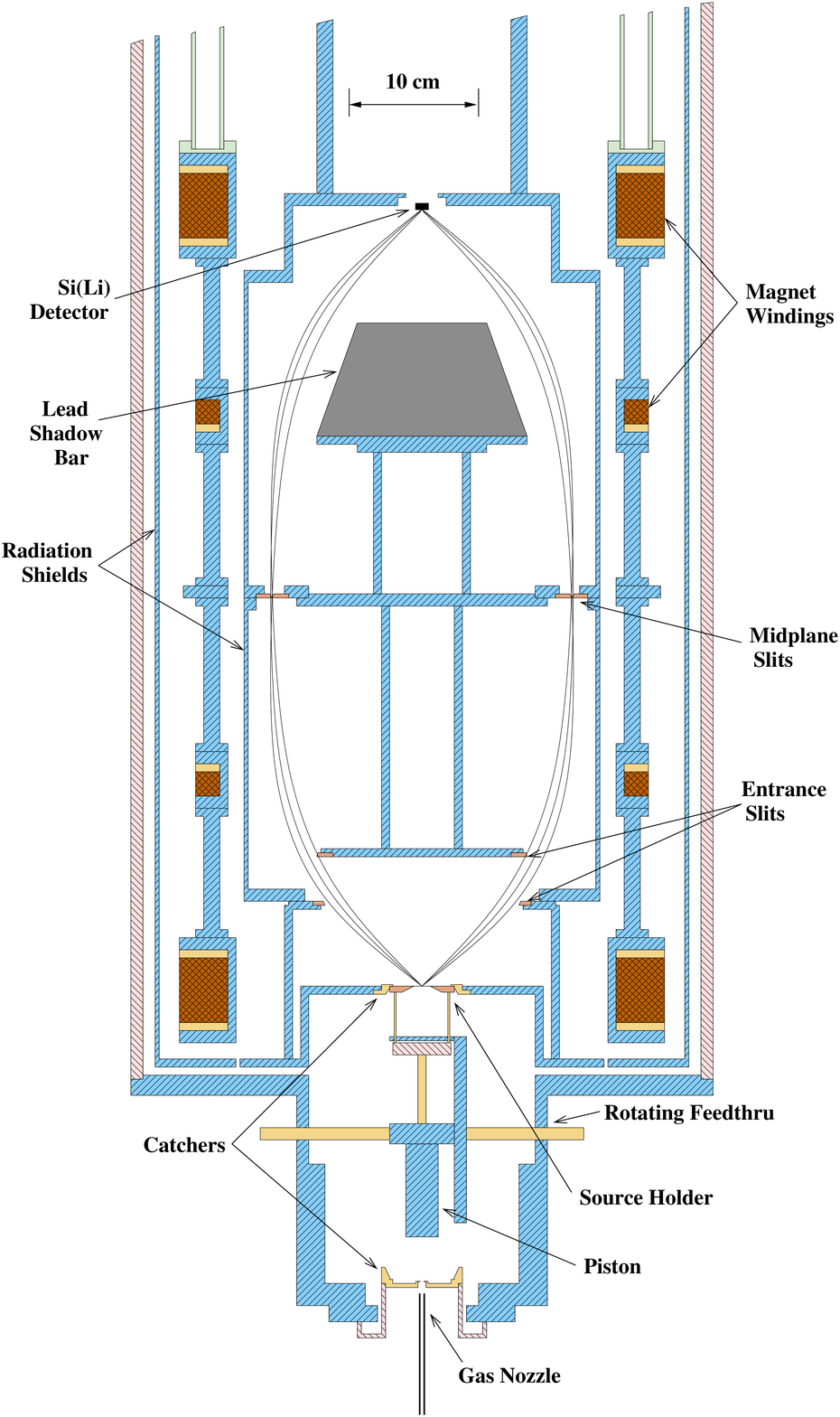}}
   \caption{(Color online) Schematic diagram of the Wisconsin superconducting 
beta spectrometer.  The apparatus is shown with
the source holder in the counting position. }
   \label{fig:spect}
\end{figure*}

\section{Experimental Details}
\label{Apparatus}
The new measurements were carried out at the University of Wisconsin Nuclear 
Physics Laboratory.  Radioactive $^{14}$O 
was produced by bombarding a
$^{14}$N gas target with a proton beam of typically 8 MeV obtained from the
Wisconsin
tandem electrostatic accelerator. A significant fraction of the 
$^{14}$O atoms were incorporated into water molecules in the production cell.
Gas from the production cell was then periodically transported through a 
capillary tube to a separation trap, 
and the H$_2$O molecules were subsequently 
transported to a beta spectrometer where the measurements
were carried out.  The details will be given below.

\subsection{Superconducting Beta Spectrometer}
The spectrometer used in the present experiment has 
been described in detail elsewhere \cite{betaspec}.
A schematic diagram that illustrates some of the relevant
details is shown  in Fig.\ \ref{fig:spect}.  
The spectrometer design follows the basic principles
of the ``Wu Spectrometer'' described in a 1956 paper by Alburger \cite {AL56},
with fields provided by a set of superconducting magnet coils.

  The magnetic fields in the spectrometer are shaped to provide
angle focusing and momentum dispersion of positrons
or electrons at the midplane.  Curves shown in Fig.\ \ref{fig:spect} 
   depict the
trajectories of positrons emitted from the source position
at angles in the neighborhood of $48^\circ$. 
These curves display 
the $r$ and $z$ coordinates of the trajectories, while in reality the
particles also spiral around the magnetic field lines.
Positrons of the appropriate momentum are focused at the
midplane slits, and after passing through the aperture are re-focused
onto a detector.

The acceptance of the spectrometer is defined by a pair of ``entrance slits''
that limit the angular acceptance, and a second pair of slits at
the midplane that
do the momentum selection.  The slits are made of 3.2\ mm thick
copper and are machined at angles so that passing positrons do not
strike the slit edge.
Since the spectrometer is iron-free,
the centroid of the momentum acceptance function
scales accurately with current. For a
current of 10\kern 2pt A, the  acceptance function peaks at approximately
2.48 MeV/c.
Under the conditions of the present experiment the acceptance function
has a FWHM of about 2\%, and a peak solid angle of roughly 0.5 sr.
The calibration of the spectrometer (momentum {\it vs.} current)
has been determined to better than
1 part in $10^4$.  The calibration procedure and many additional details
concerning properties and operation of the spectrometer are described
in Ref.\ \cite{betaspec}.

Detection of positrons that pass through the spectrometer slits
is accomplished with a nominally 1\kern 2pt cm diameter, 5\kern 2pt mm thick 
lithium-drifted silicon [Si(Li)] detector.  
Signals from the detector are processed
with a  preamplifier followed by a linear amplifier and some gating
electronics.  The signals are then analyzed with a peak-sensing 
analog-to-digital converter (ADC) which is read out by a computer.

The activity in the spectrometer is monitored with a 7.5\kern 2pt cm diameter,
5\kern 2pt cm thick BGO scintillator that is used to
detect 2.3 MeV $\gamma$-rays emitted following
$^{14}$O decay to the $^{14}$N first excited state.  
  This detector is heavily shielded against
backgrounds, and is located
at a point where it detects $\gamma$-rays that originate  near
the source position.  
The signals from this detector are
also processed through an ADC which is
read out into the computer.

\subsection{Source Preparation}

To obtain measurements of the beta spectrum we need to prepare  $^{14}$O
sources and insert them into the counting position as shown in 
Fig.\  \ref{fig:spect}.
Accurate positioning of the source is critical.  For example, a
vertical offset of 0.1\kern 2pt mm would lead to an unacceptable 
momentum shift $\Delta p\over p$ of 5 parts in $10^4$.
Furthermore, in view of the short half-life, the preparation and 
insertion of new sources obviously needs to be repeated many times.
 
In the present experiment, the source consists of
$^{14}$O water deposited and frozen onto a 3\kern 2pt mm diameter
spot at the center of a 13\kern 2pt ${\rm  \mu m}$ thick aluminum foil.
The foil is suspended across a 15\kern 2pt mm diameter hole in a copper
source holder, and attached to the source holder with epoxy.
To prevent sublimation of the water, the source holder is cooled to 
typically 140\kern 2pt K by thermal contact with ``catchers'' at 
liquid nitrogen temperature. We use aluminum because it has a greater
coefficient of thermal expansion than copper.  The consequence is that when
the source  mechanism is cooled, the foil stretches tightly across
the opening in the holder, minimizing possible longitudinal position errors.
 
The source holder can be moved between two positions.  In the lower
position, $^{14}$O water is loaded onto the foil, while the upper location
is the counting position.  In both locations the source holder is positioned by
direct mechanical contact with a catcher which centers the foil horizontally 
and fixes the vertical position.  

Since $^{14}$O has a short lifetime, the source holder is cycled between
the loading and counting positions at
frequent intervals.  For most of the measurements presented in this paper
the cycle time was 140\kern 2pt s. During a given cycle the source was
in the upper and lower positions for about 105\kern 2pt s and
25\kern 2pt s, respectively, with about 5\kern 2pt s for each transition.

To make a transition, the source holder is first 
retracted, the entire mechanism is then rotated through $180^\circ$, 
and finally the source is extended, making contact with a catcher.
A simple pneumatic gas piston is used to retract and extend the source, 
while the rotation is accomplished with a vacuum feedthrough coupled to
a stepping motor.  In the counting position the source spot is on 
the upper 
surface of the aluminum foil, so that positrons do not pass through the foil.

While the foil position is supposedly fixed by contact with the catcher, 
complete insertion of the source holder sometimes takes place slowly and 
can, at times, fail entirely.  To eliminate the resulting 
bad sections of data, the  
extension of the source holder is measured and recorded 
every 0.1\kern 2pt s.  This is accomplished with a pair of small, concentric
mutual
induction coils, one attached to the source holder and the other to the
source motion mechanism.  A sinusoidal voltage is applied to the
primary coil and the induced signal in the secondary is 
rectified, integrated and amplified, and the resulting DC signal is processed 
through an ADC.

The source motion mechanism is designed to minimize positron backscattering
by keeping the amount of material behind the source in the
acceptance cone of the spectrometer to a minimum.  In particular, 
the source holder is supported by two thin brass rods which couple 
the source to the mechanism.

Within each cycle, various operations take place at not only the 
spectrometer but also at the production cell and water separation trap.
All of the activities are automated, taking place under control of a
computer.  

\begin{figure}
\centerline{\includegraphics[width=80mm]{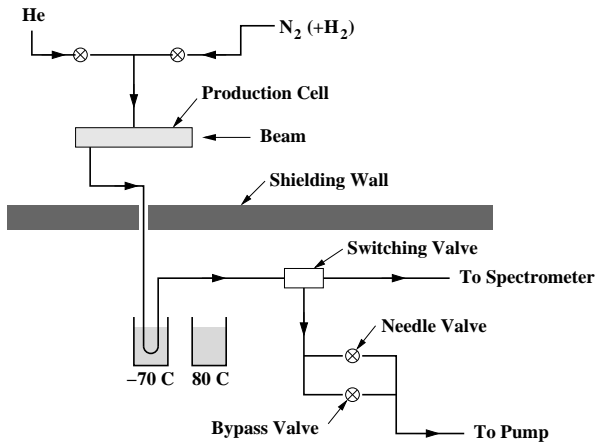}}
   \caption{Schematic diagram of the gas processing system.  $^{14}$O 
water is carried from the production cell through a thin teflon tube  and
trapped in a cold alcohol bath.  The water is periodically liberated
by moving the tube to a hot water bath, and is then transported to the
spectrometer.}
   \label{fig:gas}
\end{figure}

Figure  \ref{fig:gas} shows some of the gas handling details.
For most of each cycle, the production cell is filled at a pressure of
about 220 kPa with  nitrogen gas which has been admixed with about 
0.2\% hydrogen.  The cell volume is $\rm 3.2\kern 2pt cm^3$ with a pathlength
for beam protons of about 10\kern 2pt cm.
Within the cell much of the gas is ionized by the beam
(usually 1.0 or 1.5\kern 2pt $\mu\rm A$), and one finds 
that a significant fraction of the $^{14}$O produced gets incorporated into
water, provided that the levels of CO$_2$ and O$_2$ (which
efficiently scavenge $^{14}$O) are sufficiently low.

As $^{14}$O is being produced, gas is slowly drawn out of the cell through a 
7\kern 2pt m long teflon capillary tube with an inside diameter of 
0.8\kern 2pt mm.  Water in the gas mixture is trapped at a point where a
single loop of the teflon tube, 2 or 3 cm in length,
dips into an alcohol bath at a controlled temperature of -70 C.  Nitrogen
and other gasses that are not trapped pass through a switching
valve and a needle valve to a pump.

At some point in the cycle we wish to empty the cell and send the accumulated
$^{14}$O water to the spectrometer.  The sequence is to close
the nitrogen inlet valve at the production cell and simultaneously
open a bypass valve to increase the gas flow through the teflon tube.
After 3 seconds the cell pressure is partially reduced
and the cell is purged with a puff of helium gas followed 3
seconds later by a second puff.  After another few seconds the
condensed water is liberated by moving the teflon loop
from the cold bath to a hot water bath at +80 C.  Then, at the appropriate 
moment the switching valve is activated, routing the output gas to a 
second 0.8\kern 2pt mm diameter teflon tube that
leads to the spectrometer.  The timing parameters are all carefully 
adjusted so that there is still an adequate flow of helium from the 
production cell to sweep the water molecules along to the spectrometer.
 
The capillary tube leading to the spectrometer
was 8\kern 2pt m long for our initial data acquisition runs
and 5\kern 2pt m long in later runs.  The tube
terminates about 2 mm below the lower
source position so that the jet of helium and water is sprayed
directly onto the source foil.  Here,
a 3\ mm diameter aperture directly in
front of the foil limits the size of the active spot to that diameter.
Measurements indicate that typically $1\over 2$ to $1\over 3$ of the $^{14}$O 
activity ends up on the source foil.  Most of the remaining activity 
probably remains frozen on the collimator, which is outside the field of
view of the spectrometer.
The source foil activity was typically a few times $10^6$\kern 2pt Bq at the 
start of a counting interval.

After the $^{14}$O has been loaded onto the foil 
and most of the residual gas pumped away,
the source is moved to the upper position.  
Over time, an easily visible 3\kern 2pt mm diameter
ice spot appears on the aluminum foil.  Of course, besides the $^{14}$O
water our system also traps water that forms from oxygen that outgasses 
from the walls of the production cell.
In order to reduce backscattering of positrons from this ice,
the foil is warmed to near room temperature every few hours.  To
get some idea of how much material had collected on the foil we
combine measurements of the pressure rise as a function of time
during the warming period with an estimate of the pumping speed.
The conclusion is that the thickness of ice was generally
less than 2\kern 2pt mg/cm$^2$.



\subsection{Computer System}

A dedicated computer equipped with analog and digital I/O boards is used 
to control the experiment and collect data.  The computer performs
many jobs.  It controls the valves in the gas handling system 
and the motion of the water separation trap.  It initiates retraction,
rotation and extension of the source motion mechanism and monitors 
the resulting foil position.  It measures the current in the 
superconducting magnet and sends feedback signals to the magnet power supply
to regulate the current at the desired value.  It reads digital information
from the Si(Li) and BGO ADCs and issues the appropriate reset signals.  
Finally it provides run start and stop signals to external electronics, 
and shuts
down the superconducting magnet if temperatures drift too high.

A second dedicated computer is used to view the incoming data in
real time.  We do this to avoid the use of graphics displays on the
control computer, which create excessive dead time.
The two computers communicate through an internet link.  

The control computer also carries out the task of 
saving incoming data into an event stream.  The event stream
consists of a series of records corresponding to events of various kinds.
Recorded events include ADC outputs, measurements of the
magnet current and the foil position, plus run start and stop commands.
Each record includes an event-type identifier and a timestamp.

\subsection{Measurement Procedure}

Measurements will be reported for currents ranging from 9.5 to 18.0\kern 2pt A,
corresponding to positron momenta of 2.36-4.46\kern 2pt MeV/c.  
At lower currents one begins to encounter positrons from decay
to the $^{14}$N first excited state.  Some data were also taken at 18.5 and
19.0\kern 2pt A, currents which are near or
above the endpoint of the ground state transition.

Data acquisition was divided into a series of runs with lengths anywhere
from 20 to 45 minutes, allowing for a number of 140\kern 2pt s cycles.
Recall that our goal is to determine the shape of the
ground state beta decay spectrum, and for that purpose
we want to measure ratios of counting rates at
different spectrometer settings.  Thus within any given run
we take measurements at anywhere from 2 to 6 different magnet currents.

Once a newly prepared source has been inserted into the 
counting position we have roughly 105\kern 2pt s to observe 
decay positrons. During each 105\kern 2pt s counting period we 
cover all currents of interest for that particular run, first counting, 
then ramping to the next current, counting, ramping, and so on.
All of this is timed to complete the last current just before the
counting period ends and the source is retracted for re-loading.

The spectrometer magnets have an inductance of 12\kern 2pt H, and
consequently the ramping times are not small.  Therefore we
alternate between ``up ramps'', from low to high currents,
for one cycle, and ``down ramps'', from high to low, for the next.

Many different current combinations (or ramping modes) 
were employed in our data productions runs.  For example,
Mode 1 covers currents of 11.0, 11.5 and 12.0\kern 2pt A, while
modes 2, 3 and 4 use 6 currents separated by 1.5\kern 2pt A starting at 
9.5, 10.0, or 10.5\kern 2pt A.  Runs of this kind allow us to
cover the region of interest with the spectrum
shape fixed either by directly measured ratios, or by ratios of ratios.
In all, around 20 different ramp modes were used at one time
or another.

The data to be presented here were obtained in a series 
of 4 running periods, two in
July of 2012, and two more in February and March of 2014.  New features 
added for the 2014 runs include: 1) run-by-run monitoring of
the ``sticking fraction'', the fraction of the  $^{14}$O activity deposited
onto the source foil; 2) a more careful measurement of beam-associated
background in the Si(Li) detector; and 3) the use of a shorter
delivery tube to the spectrometer.

\subsection{Sample Spectra}

\begin{figure}
\centerline{\includegraphics[width=80mm]{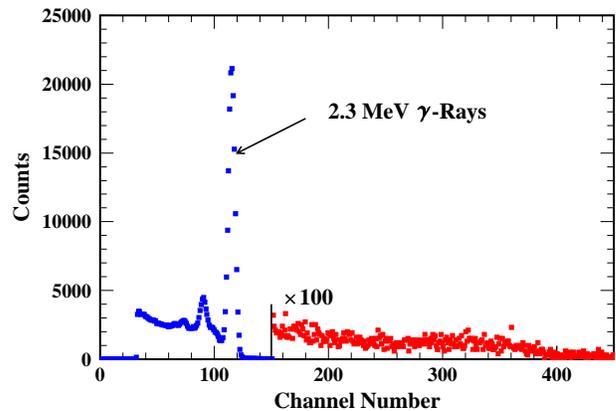}}
   \caption{(Color online) Typical energy spectrum obtained in a BGO detector
located outside the spectrometer about 1.7\kern 2pt m from the source
position.}
   \label{fig:BGO}
\end{figure}
 
As noted above, a BGO detector located outside the spectrometer 
is used to monitor
the source activity.  A typical energy spectrum obtained with this
detector (Run 6454) is shown in 
Fig.\ \ref{fig:BGO}.  Here we see a single strong peak, just above channel 100,
corresponding to 2.3 MeV $\gamma$-rays emitted 
following beta decay to the first excited state of $^{14}$N.  
Above the peak there are counts which arise primarily from beam
associated background.  Although
the BGO detector is separated from the production cell by a thick shielding
wall, some neutrons produced by the beam thermalize, diffuse into the 
spectrometer hall, and capture in various objects.  
The capture $\gamma$-rays produce background in the BGO and,
to some extent, in the Si(Li) detector as well.
In our analysis of the data (see Sect.\ 
\ref{analysis} below) we use the counting rate in this upper region of the
BGO spectrum as an aid to help us
eliminate the corresponding Si(Li) detector background.

\begin{figure}[b]
\centerline{\includegraphics[width=80mm]{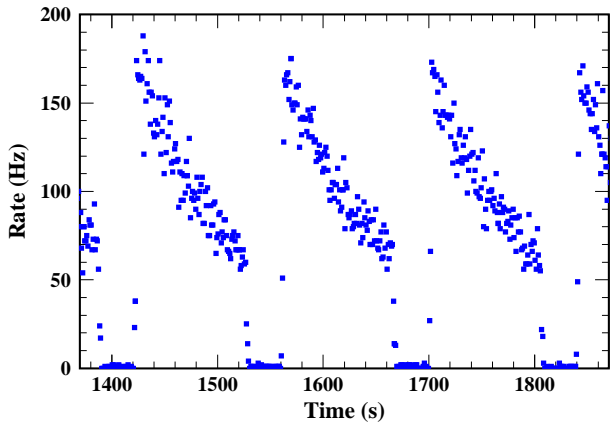}}
   \caption{(Color online) Counting rate in the BGO 2.3 MeV 
peak as a function of time.
The BGO detector is heavily shielded and mainly detects $\gamma $-rays
that originate from near the counting position.}
   \label{fig:bgorate}
\end{figure}

If we set a window around the 2.3 MeV peak, we can monitor the
source activity. In Fig.\ \ref{fig:bgorate} we show the counting
rate inside this window as a function of time for a portion of Run 6454.
We see initial counting rates between 150 and 200 Hz, with the activity
decaying as expected.  The lower source position is outside the
field of view of the detector, so the rate drops to near zero when
the source is removed for re-loading.

\begin{figure}
\centerline{\includegraphics[width=85mm]{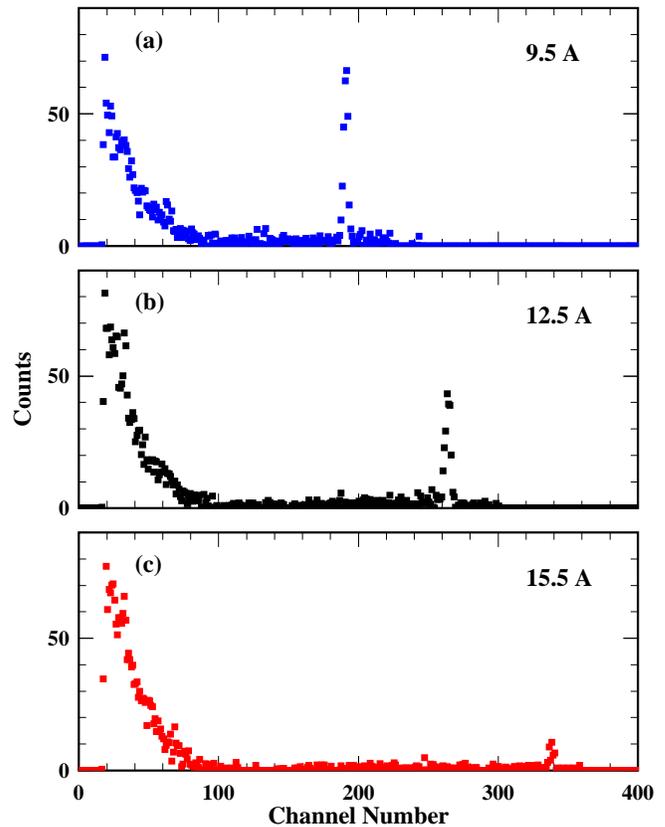}}
   \caption{(Color online) 
Representative Si(Li) energy spectra obtained during
a single run.
Results are shown for three different spectrometer
currents.  Counts below channel 100 are from positrons emitted from
$^{11}$C, while 511 keV $\gamma$-rays contribute to the rate below channel 50.
  The dispersion in the spectrum is approximately 10 keV/channel.
}
   \label{fig:silicon}
\end{figure}

In Fig.\ \ref{fig:silicon} we show some of the Si(Li)  spectra obtained
during Run 6454. These spectra have an energy gain of approximately
10\kern 2pt keV per channel.
In this particular run, data were taken at
6 currents separated by 1.5\kern 2pt A and starting at 9.5\kern 2pt A.
The plot shows the accumulated spectra for 3 of the 6 current settings,
and the main feature of interest is a peak corresponding
to the full kinetic energy of the positrons.
The run shown comprised 12 cycles and the accumulated counting 
time at each current was about 150 \kern 2pt s.  
Since we alternate
upward and downward ramps, the net number of
$^{14}$O decays at the various currents 
are the same to within about 10\%. 

The counts in Fig.\ \ref{fig:silicon} below channel number 100 are
almost entirely background, the main sources of which are 
511\kern 2pt keV gamma rays  and positrons from decay of $^{11}$C.
The presence of counts from $^{11}$C, which has an endpoint energy of
960 keV, requires some explanation.  Under the conditions of
our experiment we produce large amounts of $^{11}$C via the $(p,\alpha)$
reaction.
Most of the $^{11}$C is probably incorporated into molecules such as 
HCN, CO, CO$_2$ or CH$_4$ which would not freeze out in our water separation
trap, and should therefore be pumped away.

Now, before sending gas to the spectrometer we attempt to pump the cell and 
purge it with helium, but some  $^{11}$C certainly remains and 
can be carried to the spectrometer along with the desired remnant helium 
gas.  Any $^{11}$C deposited on the source foil
is of little consequence, since the positron momentum is far too low
for the spectrometer settings represented in Fig.\ \ref{fig:silicon}.
However, some $^{11}$C must enter the spectrometer itself.  The
teflon tube that delivers the $^{14}$O is very long and thin,
and consequently some of the $^{11}$C/helium mix continues to flow from 
the tube as the source holder begins moving to the counting position.
Any gas that emerges during the 5 s transition time is likely to enter
the spectrometer.  From there the $^{11}$C molecules diffuse 
around and are either pumped away or adsorbed onto some surface.
Based on detailed Monte-Carlo simulations, we conclude that the counts we
observe come from $^{11}$C deposited very close to the detector, possibly
even on its front surface.  That hypothesis explains the shape of the
observed low-energy Si(Li) spectrum and the fact that the rate is very
nearly independent of the spectrometer current.  

The spectra shown in Fig.\ \ref{fig:silicon} were obtained during
one of the 2012 running periods.  For the 2014 running
periods, the delivery tube to the spectrometer was
shorter and 
the $^{11}$C background was reduced by typically a factor of two or more.

\begin{figure}
\centerline{\includegraphics[width=85mm]{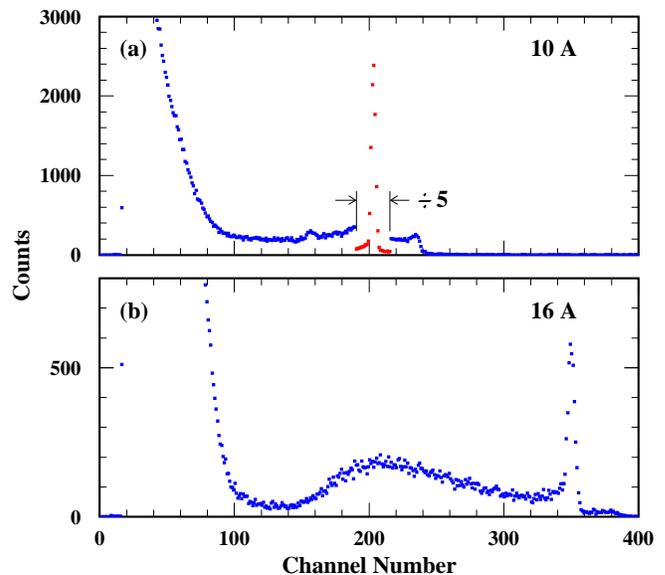}}
   \caption{(Color online)
Accumulated Si(Li) spectra for all data taken at two spectrometer
currents.  Panel (a) shows the 10 A data which correspond to about
$2.9 \!\times \! 10^{10}$ decays, while the 16 A data in panel (b)
correspond to $3.1 \!\times \! 10^{10}$ decays.
}
   \label{fig:spectra}
\end{figure}

Sample Si(Li) spectra with improved statistics are shown in 
Fig. \ref{fig:spectra}.
This figure shows the combined spectra for all data taken at currents of 10 
and 16\kern 2pt A.  In each case there is a prominent peak corresponding to
events in which the positron deposits its full kinetic energy in the detector.
The wider peak at 16 A reflects the increasing momentum acceptance of 
the spectrometer with current.

The counts above the main peak arise from processes in which the positron
energy pulse is supplemented with energy deposited in the detector
by one or both annihilation $\gamma $-rays.
Below channel 130 we see counts from $^{11}$C decay.  The broad structure
between channels 140 and 320 seen at 16\kern 2pt A is the
``$dE\over dx$'' bump, arising from
positrons that pass through the active silicon without 
depositing full kinetic energy. Finally the counts below the 10 A peak
between channels 130 and 190 
are primarily from positrons that backscatter out of the detector.

  
\section{Background Processes}
\label{bp}

Briefly stated, the goal of our data analysis 
is to determine the shape of the $^{14}$O
ground-state beta decay spectrum from 
measurements of the kind shown in Fig.\ \ref{fig:silicon}.
To accomplish this we first need to eliminate background counts.
In addition, there are various corrections that need to be applied. 
For example, there are processes that allow positrons emitted outside
the normal spectrometer acceptance to reach the detector and contribute
to the counting rate.  We begin with a discussion of the backgrounds.

Background counts can arise from a number of sources.  We have
already seen the effects of positrons from $^{11}$C
and of 511 keV $\gamma$-rays.
The counting rates from these processes are high, but fortunately the
counts are confined to the low-energy parts of the Si(Li) spectra.
At higher energies, we need to account for room background from cosmic rays 
and radioisotopes such as $^{40}$K, backgrounds associated with the beam,
and counts from 2.3\kern 2pt MeV $\gamma$-rays emitted following 
excited state decay of $^{14}$O.

\subsection{2.3 MeV Gamma-Rays}

The great majority of all $^{14}$O beta decays branch to the
$0^+$ first excited state of $^{14}$N, and are then followed by emission of 
a 2.3 MeV $\gamma$-ray.  The ground state branching ratio is only about 0.5\%
and therefore we have approximately 200 $\gamma$-rays for each
positron of interest.  The $\gamma $ flux is greatly attenuated by the
lead shadow bar located between the source and Si(Li) detector, and the
rate is further reduced by the small size and low $Z$ of the detector.
Nevertheless, 2.3 MeV $\gamma$-rays are a significant source of background.

We have no way to measure this background directly, and so we are
forced to rely on Monte-Carlo simulations.  The present simulations,
as well as others described later in this paper, were carried out
with one of two separate codes developed at 
Wittenberg and Wisconsin, both based on EGSnrc \cite {egs}.

The main process by which the 2.3 MeV $\gamma$-rays produce detector
counts in the energy range of interest
is by Compton scattering from material near the spectrometer
midplane slits (see Fig.\ \ref {fig:spect}).  When the scattering occurs near the upper surface of
a slit or the support structure, ejected electrons can follow 
field lines to the detector. 

\begin{figure}[b]
\centerline{\includegraphics[width=80mm]{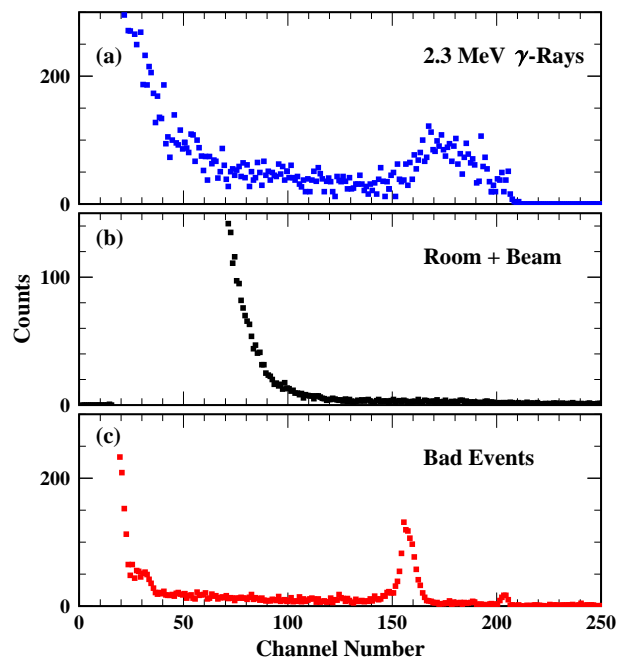}}
   \caption{(Color online)
Background sources for the Si(Li) detector at a spectrometer
current of 10\kern 2pt A.  The $\gamma$-ray and bad event spectra
are from Monte-Carlo simulations, while the room and beam backgrounds
are measured.
The backgrounds are scaled to correspond to $2.9 \times 10^{10}$ total
decays, matching the spectrum shown in panel (a) of Fig.\ \ref {fig:spectra}.}
   \label{fig:background}
\end{figure}

Results from the Monte-Carlo simulation of the 2.3 MeV $\gamma$-ray
events are shown in Fig. \ref{fig:background} along with some of
the other important background corrections.  The results presented are
for a spectrometer current of 10\kern 2pt A, and the number
of decays, $2.9\!\times \!10^{10}$, has been chosen to match the 10 A spectrum
of Fig. \ref{fig:spectra}.

In our $\gamma$-ray  simulations we see no counts above the 2.3 MeV Compton
edge, corresponding to channel 208 in our spectra.
As we increase the spectrometer current, the $^{14}$O positron peak moves up
in energy, whereas the $\gamma$ counts are still confined to the region
below the Compton edge.  We also find that the number of $\gamma$ events
decreases with increasing current. 

Besides the $^{14}$O on the source foil, many $^{14}$O atoms freeze
out in the general area of the source loading position, and of course, the 
$\gamma$-rays emitted from that location can also produce background 
counts in the Si(Li) detector.  Our simulations of this process give
spectra similar to those for $\gamma$-rays from the source position, except
that the counting rate is reduced by roughly an order of magnitude.  

Fortunately we have the possibility of doing a check on the reliability of 
the $\gamma$-ray simulations.  During data acquisition time, we took a
number of runs in which the source foil was not moved away from
the source loading position.  In this case, $^{14}$O accumulates on the
foil and the surrounding area.  From that position no positrons can reach
the Si(Li) detector, so the spectrum should be purely background.
During these ``foil-down runs'' we normally recorded the spectrum
from a second BGO detector positioned to observe 2.3 MeV 
$\gamma$-rays that originate from near the loading position.  When coupled
with Monte-Carlo calculations of the efficiency of the detector
in this geometry, we get a measure of the total $^{14}$O activity.

\begin{figure}
\centerline{\includegraphics[width=80mm]{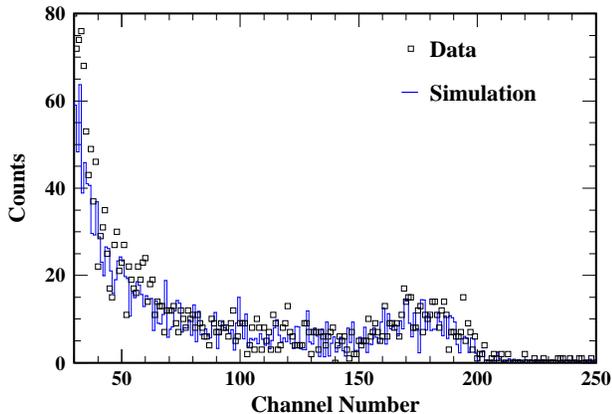}}
   \caption{(Color online)
Measurements and Monte-Carlo simulations of
the Si(Li) detector spectrum for foil-down
runs at a current of 10\kern 2pt A.  The experimental data correspond
to approximately $1.7\! \times \!10^{10}$ $^{14}$O decays, and the simulation
has been scaled accordingly. The simulation spectrum shown includes 
contributions from room and beam backgrounds.}
   \label{fig:foildown}
\end{figure}

In Fig. \ref{fig:foildown} we show the combined Si(Li) spectrum
from two such runs with the spectrometer current set at 10 A.  
From the corresponding BGO spectra we conclude that 
these runs comprised a total of about $1.7\! \times \!10^{10}$ decays.
Along with the experimental data we show the result from a Monte-Carlo
simulation of the process. The simulation spectrum shown includes
counts from the 2.3 MeV $\gamma$-rays plus
measured room and beam backgrounds (see Sect.\ \ref{OB} below).
 Given that the foil
motion mechanism is quite complex and not accurately
modeled in our simulation codes, and that the spatial distribution of the
decaying atoms is not well known, we consider the agreement between
the measurements and simulation to be surprisingly good.

\subsection{Other Backgrounds}
\label{OB}

The spectrum in panel (b) of Fig.\ \ref{fig:background}
shows the combined room and beam-associated backgrounds.
The beam-off room backgrounds were measured directly in a series of long
runs that were interspersed between the data acquisition runs.
The resulting Si(Li) rates, integrated between channels
130 and 450, are on the order of 0.015/s.  These background spectra appear to
be independent of the spectrometer current, but the overall rates 
rise very slowly during any given data acquisition period, presumably from 
activation of the surrounding materials by neutron capture.

The beam-associated background is of about this same order of magnitude.
For the 2012 data,
this spectrum was measured during a separate running period a few months
after the data acquisition periods.
During this run we accumulated Si(Li) spectra for many hours at 
different spectrometer currents with beam on and beam off.  
As in the case of the room background, it appears that the 
beam-associated background is independent of the spectrometer current.  
In 2014, the beam-on backgrounds were measured from time to time during
the data acquisition period.

It is expected that the beam background can vary from run to run,
as the beam energy, current and focusing change.  However, we believe that 
the Si(Li) beam background and the counts in the upper portions of
the BGO spectrum (above channel 200 in Fig.\ \ref{fig:BGO}) both arise from
$\gamma$-rays from
thermal neutron capture.  Therefore, the procedure we use is to compute,
for each run, the BGO rate 
summed typically between channels 225 and 450.  
We then scale the beam background spectrum by the ratio of that
rate to the corresponding BGO rate observed during the measurement of the
beam-on background.  The scale factors were typically 1 to within 10\%. 


%
%
%
  
\subsection{Corrections for Bad Events}
\label{bad_events}

We now come to the subject of backgrounds that arise from positrons.
In the simplest picture, Si(Li) counts occur when positrons 
emitted from the source pass between the spectrometer slits and reach the
detector without striking any other object. We call these ``good events''.
There are other ways in which positrons can produce 
Si(Li) counts.  Several of the mechanisms are described in detail in 
Ref.\ \cite{betaspec}.
The possibilities include scattering from slit edges, 
backscattering from the source foil, and the detection of $\gamma$-rays
from annihilation of positrons that do not reach the detector.   
These and other processes give rise to
the ``bad events''.

The good events arise from positrons emitted into a well-defined
momentum and angle window, with a momentum acceptance that scales directly
with the spectrometer current.  The spectrometer acceptance for
good events is called the ``geometrical acceptance.''  In
contrast, the bad events have no simple dependence on current.
Consequently, before making a direct comparison between counting rates
at different currents we need to remove or correct for the bad events.
As we did for the 2.3 MeV $\gamma$-rays, we determine these corrections
from Monte-Carlo simulations. 

 The simulated bad-event spectrum
at 10\kern 2pt A is shown in panel (c) of Fig. \ref{fig:background}.
Once again, the number of events has been chosen to match the 10 A spectrum
of Fig. \ref{fig:spectra}.
 The bad-event spectrum has a number of interesting features.  In particular
there is a clear peak just below channel 160.  These events arise from a 
process described previously in Ref.\ \cite{betaspec}.  Positrons emitted 
close to $85^\circ$ and with about
83\% of the nominal acceptance momentum make extra loops in the magnetic
field and can 
pass through all of the spectrometer slits.  These positrons eventually
hit the lead shadow bar (see Fig.\ \ref{fig:spect}) 
and some fraction ``bounce'' off and
reach the detector.

Ordinarily, the Si(Li) spectrum has very few of these ``loopy'' events.  
However at currents below 11\kern 2pt A, the acceptance for the loopy events
is below the endpoint for excited state positron decays, leading to a
greatly increased event rate.  For example, at 10 A, the enhancement factor is 
50, and one can see an indication of the
resulting peak just below channel 160 of the accumulated
experimental spectrum shown in panel (a) of Fig. \ref{fig:spectra}.  

Our simulations of the bad event spectrum include the effects of positrons 
that backscatter from the aluminum backing foil.  The effects are not large
in the present experiment because of the rather high positron energies.
The simulations do not account for possible spectrum distortion caused
by positron scattering and energy loss in the ice layer.  However, separate
simulations show that the ice layer effects are completely negligible.
  
\section{Data Analysis}
\label{analysis}

The data analysis proceeds on a run-by-run basis.  For each run we
have a collection of spectra similar to those shown in Fig.\ \ref{fig:silicon}.
We subtract away the room, beam and 2.3\kern 2pt MeV $\gamma$-ray backgrounds,
and then remove the bad events. Once this has been done, the spectra still
contain the backgrounds from $^{11}$C positrons and the associated 
annihilation $\gamma$-rays.  However, above about channel 130 the
spectra should contain only good $^{14}$O positron counts. 

To determine the beta spectrum we need to count all the positrons, and
since good positrons occasionally backscatter out of the Si(Li) detector, 
a significant number produce signals that lie under the $^{11}$C background.
Subtraction of this background with any degree of accuracy is not feasible.

\begin{figure}[b]
\centerline{\includegraphics[width=75mm]{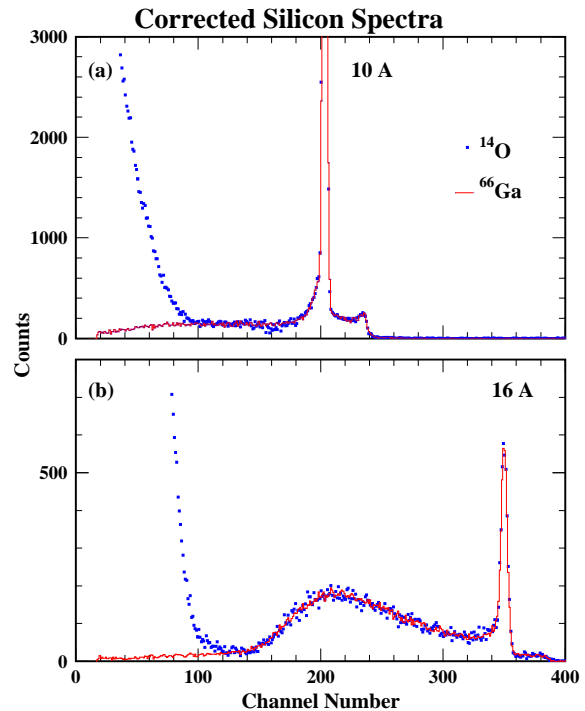}}
   \caption{(Color online)
Comparison of Si(Li) spectra obtained with $^{14}$O and $^{66}$Ga
sources.
For $^{14}$O we show the raw spectra of Fig. \ref{fig:spectra}
corrected for backgrounds and bad events.  The  $^{66}$Ga spectra
are also corrected for backgrounds and bad events, and are
normalized to match the $^{14}$O  spectrum sums above channel 130.}
   \label{fig:goodspec}
\end{figure}

Instead, we construct a ``model silicon spectrum'' for 
each current.  We then fit the model spectrum to the measurements
in the region above some threshold, and use the model to determine the 
number of good positron events below that threshold.

\subsection{Model Silicon Spectra}
\label{model}
 
In principle, one could use Monte-Carlo simulations to construct the model
spectra. We feel that simulations are fine for determining 
relatively small (typically less than 3\%) 
corrections such as the ones represented in Fig. \ref{fig:background}.  
However, the corrections for sub-threshold events 
are not small, and as we shall see, the simulations are
not adequate for determining these corrections.
Instead, we
construct the model spectra using {\bf measured} Si(Li) spectra 
from $^{66}$Ga decay.

$^{66}$Ga is a positron emitter with a half-life of 9.3 hours 
and an endpoint energy
of 4.15 MeV, just a bit higher than that of $^{14}$O.
The $^{66}$Ga was produced
by first depositing a thin layer of natural Zn onto a
 13\kern 2pt ${\rm  \mu m}$ thick aluminum foil mounted on a copper source 
holder, reproducing as closely as possible the geometry of
the $^{14}$O experiment.
The foil was bombarded with protons of about 8 MeV for a period of
a few hours, and then moved into the spectrometer counting position.
Runs were taken at all currents of interest, and room backgrounds were
measured as well.

The raw $^{66}$Ga spectra include some background events, but the situation
is much simpler than for $^{14}$O.  First and most importantly, there is no
$^{11}$C contamination.  Second, we need to account for
$\gamma$-ray induced counts, but the number of $\gamma$-rays relative
to positrons of interest is much smaller than for $^{14}$O, and all 
the $\gamma$-rays originate from the source spot.
Finally, the measurements were all taken with no beam on target.

To construct the model spectra, we carry out a series of $^{66}$Ga
Monte-Carlo simulations to determine the $\gamma$-ray backgrounds and
the bad-event spectra.  We then subtract these backgrounds along with
the measured room
background, to obtain spectra 
which should include only good positron events.

In Fig. \ref{fig:goodspec} we compare the corrected, background-
subtracted spectra for $^{14}$O and $^{66}$Ga at 10 and 16\kern 2pt A.  
For $^{14}$O we show
the combined spectra from Fig. \ref{fig:spectra} with appropriate
corrections.  The corrected $^{66}$Ga spectra have been normalized to the 
$^{14}$O result by matching the spectrum sums
above channel 130.  Except for the statistical
fluctuations the spectrum shapes match nicely down to channel 130.
Below that point the $^{14}$O spectra begin to show the effects
of $^{11}$C.

Notice that the measured $^{66}$Ga spectra are still missing some good events
below the electronic threshold at approximately channel 20. There
is also the concern that the $\gamma$-ray corrections are not small for 
the lowest energies.  Therefore, our model spectra are
constructed by using the corrected $^{66}$Ga spectra above channel 40
and Monte-Carlo simulations below that point.

\begin{figure}
\centerline{\includegraphics[width=75mm]{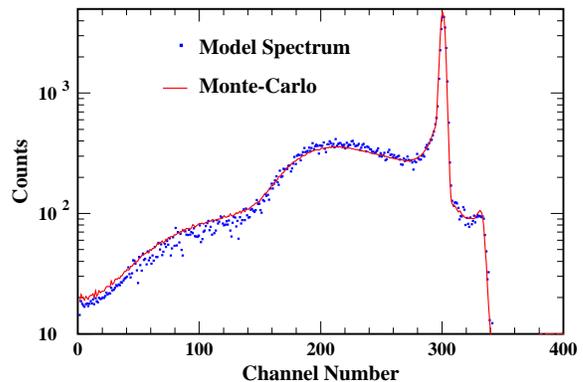}}
   \caption{(Color online)
Model Si(Li)
spectrum for a spectrometer current of 14\kern 2pt A.
The model spectrum follows the gallium measurements down to channel 40 and 
a scaled down version of the 
Monte-Carlo simulation below that point.  Note that the Monte-Carlo simulation
and the gallium measurements are not in close agreement
in the critical region around channel 100.}
   \label{fig:models}
\end{figure}

The model spectrum for a spectrometer current of
14\kern 2pt A is shown in Fig. \ref{fig:models}. Here
we also show a Monte-Carlo simulation of that spectrum, normalized in
the region above channel 130.
We can readily see that the Monte-Carlo has problems.
First, the simulation underpredicts the number of events
in the ${dE\over dx}$ bump (channels 180-270) and correspondingly overpredicts
the number in the full-energy peak.  This issue was already discussed
in Ref.\ \cite{betaspec} and has to do with uncertainties in the 
active volume of the
Si(Li) detector.  

The second problem is that  the simulation overpredicts
the number of backscattering events in the region below channel 130.
This is precisely the region where we require our model spectrum to
be reliable.  The overprediction of the backscattering occurs at
all spectrometer currents, and the fractional excess is around 10 to 15\%
from threshold up to at least channel 100.
Thus, in constructing the model spectrum we use Monte-Carlo
results that are scaled down by this amount for the region below channel 40.

\subsection{Data Extraction}
\label{model}
 
The data analysis proceeds as follows.  We begin by choosing cuts for
summation of the Si(Li) data at each energy. The cuts are chosen
to exclude all $^{11}$C events and to minimize other
potential problems.  For example, to
the extent possible we try to avoid
contributions from the 2.3 MeV $\gamma$-rays.  Similarly, as one can see in
Panel (a) of Fig.\ \ref{fig:goodspec}, the elimination of the
peak seen in the bad-event spectrum (near channel 160 in
Fig.\ \ref{fig:background})
is not perfect, 
so we choose the lower cut to be above this region for currents
less than 11\kern 2pt A where the loopy events could be problematic.
Once the cuts are chosen, we use the model spectrum to calculate
the correction factor, $F_S$, for good events outside the window.

Then, for each run, we sum the accumulated Si(Li) spectra 
(see  Fig.\ \ref{fig:silicon}) for each current of interest.  For reasons
to be explained shortly, separate sums are computed for
upward and downward ramps.  For each current and ramp direction we also
compute the required subtractions for background and bad events (see
Fig. \ref{fig:background}).

The final required quantity is a decay factor, $F_D$, defined as the fraction
of all $^{14}$O decays that occur while counting at a specific current.
Denoting the measured spectrum sums by $S_R$ and the required subtractions
by $B$, the corrected event sum $N_R(I)$ for a given run and 
a given current is then
\begin{equation}
N_R(I) = (S_R-B)F_S/F_D,
\label{eq:NI}
\end{equation}
where the subscript $R$ is to be thought of as a run number index.
Basically, Eq.\ (\ref{eq:NI}) corrects for
backgrounds and sub-threshold events, and extrapolates
the measured sums back to a common start time.

The alert reader will undoubtedly recall that each run is divided
into many separate cycles.  Within a given cycle, suppose the counting
for a particular current begins at time $t_1$ and ends at time $t_2$,
where $t=0$ is taken to be the cycle start time.
Then the decay factor is given by $F_D = e^{-t_1/\tau} - e^{-t_2/\tau}$.
These decay factors could in principle differ
from one cycle to the next.  In practice, however, the
decay factors are almost identical for all up ramps, and similarly almost
identical for all down ramps.  Consequently, we
take $F_D$ in Eq.\ (\ref{eq:NI}) to be the individual cycle
decay factors averaged over all up or all down ramps as appropriate.
Similarly, $S_R$ in Eq.\ (\ref{eq:NI}) is taken to be the number 
of Si(Li) counts (within the cut window)
summed over all up or all down ramps.

The corrected event sums, $N_R(I)$, defined in Eq.\ (\ref{eq:NI}) 
depend on the net
$t=0$ source activities, $A$, which vary from run to run but are 
the same for all currents within
a given run, and also on the $^{14}$O beta spectrum intensity averaged over
the acceptance function of the spectrometer.  This latter factor 
which we shall denote as $\bar n (I)$ depends
only on the current setting.  Thus we have
\begin{equation}
N_R(I) = A_R \ \bar n (I).
\label{eq:counts}
\end{equation}
We now need to extract the $\beta$ spectrum, 
$n(p) \equiv {dn\over dp}$, from these measurements.

\subsection{Results}
\label{results}
 
Information on the momentum dependence of the
$\bar n$'s can be obtained by 
forming ratios of the corrected event sums
to cancel the unknown $A_R$ factors in Eq.\ (\ref{eq:counts}).
In practice, however, we use a somewhat more complex procedure that optimizes
the statistical impact of each measurement.

First we fix the value of $\bar n$ at one current, 11\kern 2pt A.
We then treat the values at other currents as free parameters which are
adjusted to provide the best overall agreement with the full data set.
Here, the full data set consists of 255 runs and a total of 
2212 event sums.  
The free parameters in the fit are the 
$\bar n$'s along with the activity factors, $A_R$, of Eq.\ (\ref{eq:counts}).
Since we analyze up and down ramps separately, there are two $A_R$ values
for each run, for a total of 528 fitting parameters.
Fortunately, the best-fit $A_R$ values for any proposed set of $\bar n$'s can 
be computed algebraically.

Frequently, the individual spectrum sums are small numbers, and therefore 
we use Poisson statistics, which requires fitting the directly measured 
(integer) spectrum sums.
 By combining Eqs.\ (\ref{eq:NI})
and (\ref{eq:counts}) we isolate these quantities,
\begin{equation}
S_R(I) = A_R \kern 2pt \bar n(I)\kern 1.5pt F_D/F_S + B.
\label{eq:sums}
\end{equation}
In this equation $B$, $F_D$ and $F_S$ are all known, and the
fitting parameters are adjusted to maximize the 
Poisson likelihood function, $\cal L$.

After extracting the $\bar n(I)$ values by the
procedure described above, we need to convert to $n(p)$.
Let $\Omega _I(p)$ represent the acceptance solid angle of the spectrometer
as a function of momentum at some current $I$.  Then the observed number of
counts at that current should go as
\begin{equation}
\bar n(I) = C\kern 1pt \int n(p) \Omega _I(p) dp,
\label{eq:folding}
\end{equation}
where $C$ is some constant.
We shall make use of the fact that $n(p)$ is a smooth function 
while $\Omega _I(p)$ is sharply peaked with a centroid at 
$p_0 = r_0 \kern 2pt I $ where $r_0 \simeq 248 $ keV/A.
Also, recall that the acceptance of our spectrometer scales with
current, meaning that $\Omega _I(p) = \Omega_0({p/I})$ where
$\Omega _0$ is a universal function with centroid $r_0$.

Upon making the change of variable $ r = p/I$ 
Eq.\ (\ref{eq:folding}) becomes
\begin{equation}
\bar n(I) = C\kern 2pt I\int n(rI) \kern 2pt\Omega _0(r)\kern 2pt dr.
\label{eq4}
\end{equation}
Now if $n(p)$ is sufficiently smooth we can approximate Eq.\ (\ref{eq4})
as
\begin{equation}
\bar n(I) \simeq C\kern 2pt I\kern 2pt n(p_0) \int \Omega _0(r)\kern 2pt dr,
\label{eq5}
\end{equation}
which, with some level of error, would allow us to extract the desired 
beta intensities $n(p)$ from the measured rates $\bar n(I)$.

Through most of our energy range the approximation of Eq.\ (\ref{eq5}) 
is quite good.  In first order the rate at current $I$ just depends on the 
beta intensity at the corresponding central momentum.  In general,
however, there will be corrections
that arise from the curvature of the beta spectrum together with the finite
width of the acceptance function.  Higher order terms
in a Taylor expansion of $n(p)$ may also be relevant.

Now as one approaches the
endpoint of the beta spectrum, $n(p_0)$ tends towards zero
and the higher order terms become significant.  
For example, at 18.5\kern 2pt A the contribution from the quadratic term 
is actually larger than the  leading term.

Since Eq.\ (\ref{eq5}) is not always adequate, we apply a correction.
The correction factor is found by postulating a theoretical beta spectrum
and using that spectrum to compute the right-hand sides of Eqs.\ (\ref{eq4})
and (\ref{eq5}).  The ratio 
$\int n(rI) \kern 2pt\Omega _0(r)\kern 1.5pt dr/
[n(p_0)\kern -1pt \int\Omega_0(r)\kern 1.5ptdr]$
then becomes our
correction factor.  This procedure should be fine 
for $^{14}$O decay since the spectrum does not deviate greatly from the allowed 
shape.  We find that the correction is a factor of 3.7 at 18.5 A, 
but only 3\% at 18.0 A 
(enough to move that point by one error bar) and less than 1\% at lower
currents.  This same  procedure was used in Ref.\ \cite{gallium_shape}
with excellent results.

\begin{figure}[b]
\centerline{\includegraphics[width=75mm]{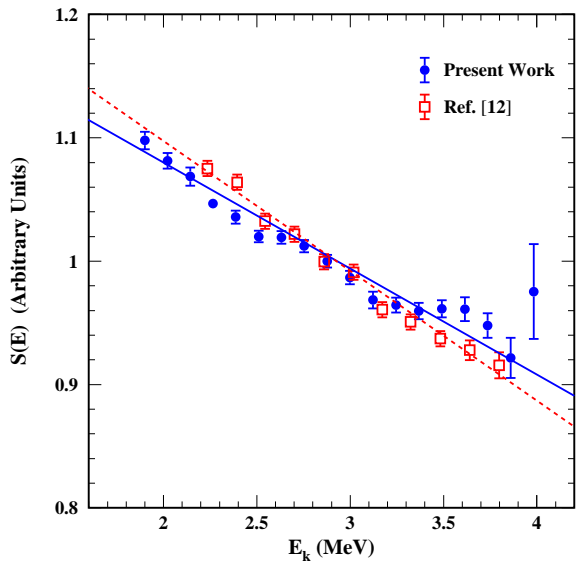}}
   \caption{(Color online)
Measurements of $S(E)$ from Ref.\ \cite{sidhu} and from the
present work, plotted as a function of the
positron kinetic energy.  Both data sets have been normalized to
unity at roughly the center of the measured energy range. Recall that
the uncertainties in the new measurements are strongly correlated.
The straight lines are guides to the eye.}
   \label{fig:dataplot}
\end{figure}

Recall that in our experiment, we determine only ratios 
of the $\bar n(I)$ values at different currents and that we have 
 fixed the value of $\bar n$ at 11\kern 2pt A.  
In the plot to follow, the corresponding $n(p)$ value is shown
without an error bar.  The uncertainties shown for the remaining data
points are effective 1$\sigma$ Poisson errors, obtained by locating the 
points in parameter space where $\ln {\cal L}$ 
is smaller than the maximum value by 0.5.  Because
we measure ratios and ratios of ratios, the uncertainties
in the $n(p)$ values are strongly correlated.  

Our results are presented in Fig.\ \ref{fig:dataplot}.  Here we plot the 
ratio of the extracted spectrum, $n(p)$, to the allowed statistical shape:
\begin{equation}
S(E) = {n(p)\over p^2\kern 2pt(E_0 - E)^2\kern 2pt F_0(p,Z)},
\label{eq:shape}
\end{equation}
where $E$ is the positron energy, $E_0$ is the endpoint energy corrected
for nuclear recoil (4.63223 MeV),
and $F_0$ is the usual Fermi function for a point charge nucleus
with lepton wave functions evaluated at the nuclear surface (see for example
Ref.\ \cite{wilk1}).  The data set has been normalized to
$S(E) = 1$ at roughly the center of the measured energy range.
For comparison we also show the
previous measurements of this quantity reported by SG 
\cite{sidhu}, normalized in the same way.
We see that the two data sets are similar, although the new measurements
have a somewhat smaller slope than the previous ones.  Also, the new
measurements extend over a larger energy range.

Besides the points shown
in Fig.\ \ref{fig:dataplot} we have measurements at both 18.5 and 19.0 A.
Determination of $n(p)$ at the highest currents is complicated
by the low counting rates for real positrons and by
the proximity to the endpoint which leads to large corrections for
the finite acceptance of the spectrometer.  
At 18.5\kern 2pt A, where the kinetic energy at the centroid of the
acceptance function is only 10\ keV below the endpoint,
we observe a total of 208 counts with an expected background of 179.
After applying the finite acceptance correction factor 
of 3.7, the resulting $S(E)$ measurement is $0.7 \pm 0.4$ in 
agreement with the trend of
the data shown in Fig.\ \ref{fig:dataplot}. Our highest current
(19 A) is above the $^{14}$O endpoint.  There we have
101 observed counts with an expected background of 104.  These
high-current results give us confidence that our treatment
of the backgrounds and the finite acceptance correction are adequate.

\section {Determination of the Shape Parameters}

Before extracting detailed shape information from our data, we want to
account for all the energy dependences that are known to be present
for purely allowed transitions.
For this purpose we define a modified Fermi function
\begin{equation}
F(p,Z) = F_0(p,Z)\kern 2pt L_0^A \kern 2pt C_A\kern 2pt R_A\kern 2pt Q\kern 2pt
g(E,E_0).
\label{eq:fermi}
\end{equation}
Here $g(E,E_0)$ is a radiative correction factor
calculated following Sirlin \cite{sirlin}, $L_0^A$ and $C_A$ are finite size
corrections as given by Wilkinson \cite{wilk1}, and $R_A$ 
and $Q$ are corrections for recoil and screening, respectively, 
again calculated according to Wilkinson \cite{wilk2}.

We then construct a shape function with this modified Fermi function:
\begin{equation}
C(E) = {n(p)\over p^2\kern 2pt(E_0 - E)^2\kern 2pt F(p,Z)}.
\label{eq:shape2}
\end{equation}
Most of the correction factors that appear in Eq.\ (\ref{eq:fermi})
are quite close to 1, the exception being the radiative correction
term.  That factor has a negative slope and varies by about 2\% over
the range of our measurements.  Consequently, the plot of $C(E)$ looks much
like $S(E)$ in Fig.\ \ref{fig:dataplot}  except with a slightly reduced slope.

Based on past theoretical work (see Sec.\ \ref {History}), we expect 
that contributions from terms beyond $\langle \sigma \rangle$ and
$\langle {\rm WM}\rangle$ are not negligible, 
and therefore it is not appropriate to
fit our $C(E)$ measurements with a function of the form given in
Eq.\ (\ref{eq:slope}).  Instead we exploit the observation
by Behrens and B\"uhring (see Ref.\ \cite{BB}, page 462) 
that the shape function for allowed transitions can be written in the form
\begin{equation}
C(E) \simeq k(1 + aW + b/W + c W^2),
\label{eq:shapefunction}
\end{equation}
where $W$ is the positron total energy in units of its rest energy, 
and where $k$ and the shape parameters $a$, $b$ and $c$ are all constants.

\begin{figure}[b]
\centerline{\includegraphics[width=75mm]{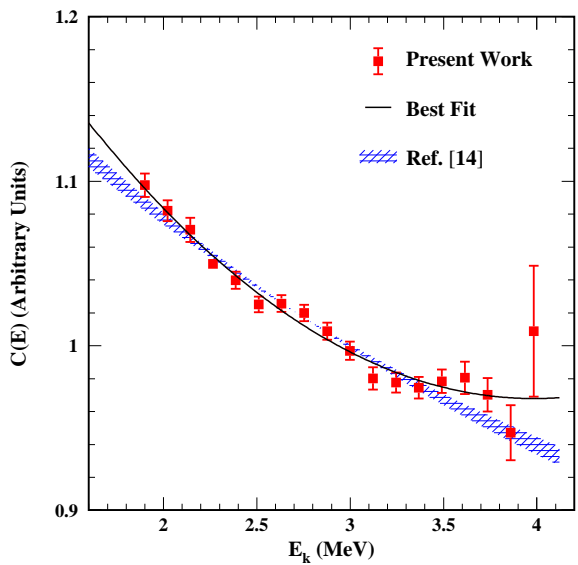}}
   \caption{(Color online)
Experimental values of $C(E)$  along with the best fit to 
the current data set. The experimental points and the solid curve shown are
obtained separately from Poisson fits to the run-by-run measured counts. 
The hashed region shows the range of the 
theoretical $C(E)$ predictions from TH \cite{TH}. The
curves terminate at the endpoint of the transition.
Recall that the uncertainties in the data points are strongly
correlated.  }
   \label{fig:C_E}
\end{figure}

Since we have measurements of $C(E)$ over a limited energy range, it is 
completely impractical to determine all three shape parameters, and so to 
develop a fitting strategy,
we turn to the theoretical calculations for guidance.    Towner
and Hardy \cite{TH} tabulate
the shape parameters for a variety of shell model wave functions, while
Garc{\'i}a and Brown 
\cite{GB} provide enough information to allow these quantities to be
computed.  All the calculations we have seen suggest that the $b$ term
is very small over our energy range.  In addition, the $a$ and $b$ terms
are strongly correlated in such a way that a change in the  value of $b$ can
be compensated by a much smaller change in $a$.  Finally, since the calculated
$b$'s all tend to be of about the same size,
we will fix this parameter at a typical theoretical value.

If one fits measurements with $a$ and $c$, one learns that these
two parameters are also strongly correlated.  The $c$
term allows for curvature in $C(E)$, but if this term becomes substantial,
it contributes to the overall slope.  In this way, uncertainty
in the curvature translates into an uncertainty in $a$.
To get around this difficulty, we fit the data with an algebraically 
equivalent formula,
\begin {equation}
C(E) = k'[1 + a'W + b'/W + c' (W -  W_c)^2],
\label{eq:shapefunction}
\end{equation}
where $W_c$ is taken to be the $W$ value corresponding to a kinetic
energy of 2.75\kern 2pt MeV, close to the middle of our
measured energy range.
With this choice the $c'$ term has zero slope at $W_c$,
virtually eliminating the correlation between $c'$ and $a'$.
 
The fitting procedure basically follows the method outlined in
Sec.\ \ref {results}. Given  $a'$, $b'$ and $c'$, we compute 
$\bar n (I)$ from Eq.\ (\ref{eq:folding}) and then fit
the measured sums via Eq.\ (\ref{eq:sums}) by maximizing
the  Poisson likelihood function. 
With $b'$ fixed at 0.04 we obtain
the central result of the present work:
\begin{eqnarray}
&& a' = -0.0290\pm 0.0008, \nonumber \\
&& c' = \kern 2.8mm 0.0061 \pm 0.0010.
\label{eq:result}
\end{eqnarray}
In Eq.\ (\ref{eq:result}) the quoted uncertainties are 
once again the effective 1$\sigma$ 
statistical Poisson errors.  
The quality of the fit is good.  A detailed statistical analysis indicates
that for the current problem one should expect best fit $\ln \cal L$
values in the range $-9040 \pm 30$, and our result is $\ln {\cal L} = -9025$.

Given the primed shape parameters
it is straightforward to calculate the unprimed ones.  The result is
$a = -0.0865$, $b = 0.032$, and $c = 0.0050$.  Our fixed value for
$b'$ was chosen to give $b$ in agreement with typical
theoretical predictions (see for example Ref.\ \cite{TH} Table III).


The best fit is shown in Fig.\ \ref{fig:C_E}, along with our experimental 
results for $C(E)$.  The agreement between the curve
and the experimental points seems to be good, though we need to remember that
the experimental uncertainties are strongly correlated.  Nevertheless, 
it is clear from the plot that the measurements
favor a fit with a positive curvature.

\subsection {Systematic Errors}

There are several non-trivial sources of systematic error in
the shape parameters.
We use measured $^{66}$Ga spectra and Monte-Carlo calculations to correct
for positron events that fall below the threshold of our
Si(Li) summation window.  The $^{66}$Ga measurements have statistical
uncertainties, and we also found it necessary to renormalize 
the Monte-Carlo spectra in the region below channel 40.
We estimate that the combined uncertainties associated with the sub-threshold
corrections are $\delta a' = 0.00041$ and  $\delta c' = 0.00037$.

The spectrometer is calibrated to an accuracy of 1 part in $10^4$
leading to an uncertainty in the central momentum of the detected positrons.
The resulting systematic error contributions are
$\delta a' = 0.00025$ and  $\delta c' = 0.00018$.  We apply
corrections for flux pinning in the superconducting magnets 
\cite {betaspec}.  Taking the uncertainty to be half the correction
gives $\delta a' = 0.00007$ and  $\delta c' = 0.00001$.

We subtract counts that result from 2.3 MeV $\gamma$-rays, but there
are possible systematic errors in the required Monte-Carlo simulations.
We estimate the resulting systematic uncertainties to be
$\delta a' = 0.00034$ and  $\delta c' = 0.00029$.
Uncertainties arising from the other background sources are small,
$\delta a' = 0.00009$ and  $\delta c' = 0.00010$.
The uncertainties associated with backscattering from the aluminum
source foil are similarly small, 
$\delta a' = 0.00003$ and  $\delta c' = 0.00002$.

Besides the error sources listed above, we have also considered
systematic errors arising from the uncertainties in the
$\beta$-decay $Q$ value and lifetime, from deadtime and pileup, 
and from the uncertainty in the spectrometer acceptance width
(see Ref.\ \cite{gallium_shape}).
All of the resulting 
systematic error estimates are negligible.

Combining all of the systematic uncertainties in quadrature, we obtain net
systematic errors of
\begin{equation}
\delta a' = 0.0006 \kern 2mm {\rm and}\kern 2mm  \delta c' = 0.0005.
\end{equation}

\section {Discussion}

Let us compare our shape parameters with those obtained from the 
measurements of SG \cite{sidhu}.  For this purpose we use the
$C(E)$ values and uncertainties given in Table 1 of TH \cite {TH}.
We fit these data with the formula given in
Eq.\ (\ref{eq:shapefunction}), taking $k'$, $a'$ and $c'$ as
free parameters and fixing $b'$ at 0.04.  The results are
\begin{equation}
a' = -0.0390\pm 0.0017 \kern 2mm {\rm and}\kern 2mm 
c' = 0.0044 \pm 0.0019.
\end{equation}
As we expect from Fig.\ \ref{fig:dataplot}, the SG data have a greater
slope than the new measurements. Somewhat unexpected is the
fact that the SG data also favor a positive curvature, statistically
consistent with our result.  One
can possibly see a slight curvature in data of Fig.\ \ref{fig:dataplot},
but that curvature is enhanced by the correction factors which are applied to
convert $S(E)$ to $C(E)$.

The central question is whether the new measurements are
consistent with CVC.
In the naive picture in which
GT and WM are the only non-zero matrix elements, $C(E)$ should be
of the form Eq.\ (\ref{eq:cvc}) with
\begin{equation}
{4\over 3M} {\langle {\rm WM}\rangle\over \langle\sigma\rangle} = 
{0.0248\over m_e},
\label{eq:cvcvalue}
\end{equation}
provided that CVC and charge symmetry hold. This conclusion is correct
whether or not one renormalizes the GT and M1
(or equivalently WM) operators as described in Ref.\ \cite{TH}.
Our measurements have a greater 
slope than one obtains from Eq.\ (\ref{eq:cvcvalue}), and
as we have emphasized, the measurements also have a non-zero
curvature which is inconsistent with Eq.\ (\ref{eq:cvc})
no matter what the value of $\langle {\rm WM}\rangle$. Therefore, higher 
order terms are clearly required.

At this point it is helpful to turn to detailed shell model calculations.
Towner and Hardy \cite{TH} have reported shape parameters for a number of
such calculations.  These authors begin with 
shell model wave functions from various sources and then allow mixing 
between the first and second $1^+$ eigenvectors of the calculation,
by fitting the $C(E)$ measurements of SG.  This is almost the
same as fitting the $ft$ value. When the authors
use free particle operators, the best fit theoretical $C(E)$
has about the right slope, but the resulting $\langle {\rm WM}\rangle$
values are not consistent with CVC and the known M1 electromagnetic decay.
The calculations are then repeated with renormalized GT and WM operators,
and in this case the $\langle {\rm WM}\rangle$ values are
within 10\% of the CVC value. 

Results obtained with renormalized operators are given in Table III of 
TH \cite{TH}.  These calculations \cite {PC} all have $a'$ values  
in the range $-0.0276$ to $-0.0308$, covering our
measured value.  
The ``CK'' calculation, for which $\langle {\rm WM}\rangle$ matches
the CVC prediction to better than $1\over 2$\%, gives $a' = -0.0285$.

More can be learned from Garc{\'i}a and Brown \cite{GB}.
In Table VIII GB report values of 5 matrix elements,
$V_1^A$, $V_1^V$, $V_2$, $V_3$ and $V_4$. 
$\langle {\rm WM}\rangle$ is a linear combination of 
$V_1^V$ and $V_3$,
\begin{equation}
\langle {\rm WM}\rangle = {(\mu_p - \mu_n)\over \sqrt{2}}\kern 1.5pt 
V_1^V + V_3,
\end{equation} and  the authors choose $V_3$ to satisfy CVC. The
value of $V_1^A$, which is our $\langle {\sigma}\rangle$, is chosen to
reproduce the measured $ft$ value.
We know that the shape parameters are very sensitive to
variations in $V_1^A$, and therefore we re-optimize
this parameter to obtain a decay constant for the transition
of $\lambda = 5.3\times 10^{-5}$/s.
This result corresponds to a ground-state branching ratio
of 0.54\% obtained by TH in their reanalysis of the SG data set.

We take numerical values for the matrix elements
from Table VIII of GB, and after making
the small adjustment in $V_1^A$ we compute the shape parameters.
The resulting $a'$ values for the three shell model wave functions
considered by GB fall in the range $-0.0256$ to $-0.0281$.

\begin{figure}[b]
\centerline{\includegraphics[width=75mm]{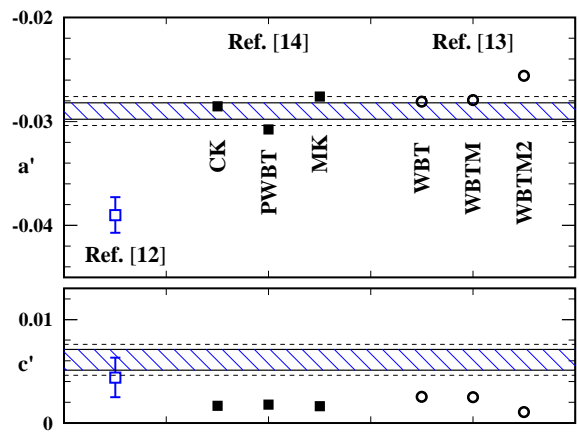}}
   \caption{(Color online)
Experimental and theoretical results for the shape 
parameters $a'$ and $c'$.  The results from the present work are
represented by the hashed bands which show the limits of the
1$\sigma$ statistical errors.  The dashed lines above and below
the bands show the sum of the statistical and systematic errors.
The $a'$ and $c'$ values we extract from the SG data set are
shown by the open squares.  Theoretical results shown
are from Table III of Ref.\ \cite{TH},
represented by the filled squares, and Table VIII of Ref.\ 
\cite{GB}, shown by open circles.
}
\label{params}
\end{figure}

Many of the results quoted above are summarized in Fig.\ \ref{params}.
In panel (a) we show our measured value of $a'$ along with the
result we extract from the SG data.  The points shown without error bars
are the TH and GB calculations discussed above, and as
we can see, the theoretical points cluster around our measured value.

The agreement between theory and experiment for the curvature
parameter is not so good.  In panel (b) of Fig.\ \ref{params} we
see theoretical $c'$ values ranging from 0.0010 to 0.0025, several
standard deviations from the measured value of 0.0061.

To understand what all of this means we would like to know 
how various matrix elements affect the calculated shape parameters.  
Suppose we start
from one of the GB calculations and arbitrarily change 
$\langle {\rm WM}\rangle$ from the
CVC value by 10\% by shifting either $V_1^V$ or $V_3$.
The result we find is that $a'$ shifts by 8\% (0.0021) while $c'$ moves
by only $5\times 10^{-5}$.  We conclude that 
the discrepancy between our measured curvature and the theoretical 
predictions has nothing to do with CVC. 

We can also investigate what happens when we shift values of the higher
order matrix elements
$V_2$ and $V_4$.  For both of these we
find that a shift in the matrix element value 
moves $a'$ and $c'$ by similar amounts.  In view of this, it seems
plausible that with the right values of $V_2$ and $V_4$, one
might improve the agreement with $c'$ without seriously degrading
the present slope parameter agreement.

Overall, we are very pleased with the results.  We
have seen that the slope parameter $a'$ is very sensitive to the value 
of $\langle {\rm WM}\rangle$, and the agreement of our 
measured value with calculations that respect CVC is excellent.
The lack of similarly close agreement with the measured curvature
parameter should probably not be a major concern.  In fact, if we
add the statistical and systematic uncertainties, the 
largest theoretical $c'$ values are less than 2$1\over 2$ standard
deviations from the measured value.

All of the theoretical $C(E)$ curves actually reproduce the new measurements
fairly well. This is illustrated in Fig.\ \ref{fig:C_E} where
the shaded band shows the range of predictions covered by the
three calculations listed in Table III of TH.  Since we have not reported
a measurement of the absolute magnitude of $C(E)$, we renormalize
the TH curves to match the scale of the plotted points.
Given that these calculations were originally optimized in an
effort to reproduce the SG data, the agreement with the present
measurements is remarkably good. The calculations of GB would
produce a similar, but slightly wider, band.

\section {Summary}
 
We have carried out an experiment designed to test CVC by measuring the shape
of the $\beta$-spectrum for the $0^+ \rightarrow 1^+$ decay of $^{14}$O to
the ground state of $^{14}$N.  The measured shape function has
a slope somewhat smaller in magnitude than that of the 
measurements reported in Ref.\ \cite{sidhu}.  The new measurements
allow us to determine the value of a parameter, $a'$,  which is 
essentially the average slope of the shape function over the energy range of
our measurements, to a relative accuracy of better than 3\%.

The measured slope parameter is in good agreement with predictions
from theoretical calculations that respect CVC by
requiring agreement between the $\beta$-decay weak
magnetism matrix element, $\langle {\rm WM}\rangle$, and the M1 matrix 
element for the electromagnetic decay of the first excited state
of $^{14}$N.

The authors wish to thank Alejandro Garc{\'i}a for suggesting
the experiment. We also thank Laura Kinnaman, Megan McGowan and Li Zhan
for their assistance on the project, and the University of Wisconsin 
Center for High Throughput Computing for allowing access to their
computational facilities.
This work was supported in part by the National Science
Foundation under grants Nos. PHY-0855514 and PHY-0555649, and in part by
an allocation of time from the Ohio Supercomputer Center.


\end{document}